\documentclass{iaus}
\usepackage{graphicx}

\title[Radiation from matter entrainment in astrophysical jets] 
{Radiation from matter entrainment in astrophysical jets: the AGN case}

\author[Araudo, Bosch-Ramon \& Romero]   
{A.T. Araudo$^{1,2}$, V. Bosch-Ramon$^{3}$  
 \and G.E. Romero$^{1,2}$}

\affiliation{$^1$
Instituto Argentino de Radioastronom{\'\i}a (CCT La Plata, CONICET),\\ 
C.C.5, 1894 Villa Elisa,  Buenos Aires, Argentina,\\
email: {\tt aaraudo@fcaglp.unlp.edu.ar, romero@fcaglp.unlp.edu.ar} \\
[\affilskip]
$^2$ Facultad de Ciencias Astron\'omicas y Geof\'{\i}sicas,
Universidad Nacional de La Plata,\\ Paseo del Bosque, 1900 La Plata,
Argentina\\[\affilskip]
$^3$ Dublin Institute for Advanced Studies, 31 Fitzwilliam Place,
Dublin 2, Ireland \\email: {\tt valenti@cp.dias.ie}}

\pubyear{2010}
\volume{275}  
\pagerange{119--126}
\jname{Title of your IAU Symposium}
\editors{A.C. Editor, B.D. Editor \& C.E. Editor, eds.}
\begin{document}

\maketitle

\begin{abstract}
Jets are found in a variety of astrophysical sources, 
from young stellar objects to active galactic nuclei. 
In all the cases the jet propagates
with a supersonic velocity through the external medium, which can be 
inhomogeneous, and inhomogeneities could penetrate into the jet.  
The interaction of the jet material with an obstacle
produces a bow shock in the jet in which particles can be accelerated up to 
relativistic energies and emit high-energy photons. 
In this work, we explore the active galactic nuclei scenario,
focusing on the dynamical and radiative consequences of the interaction
at different jet heights. 
We find that the produced high-energy emission could be
detectable by the current $\gamma$-ray telescopes. In general,  
the jet-clump interactions are a possible mechanism
to produce (steady or flaring) high-energy emission in
many astrophysical sources in which jets are present.

\end{abstract}

\firstsection 
\section{Introduction}

Jets at different scales are present in astrophysical sources such as 
 protostars, microquasars (MQ) and  active
galactic nuclei (AGN). The medium that surrounds the jets is not homogeneus
and clumps from the external medium can interact with them. 
In young stellar objects (YSO), the interaction of matter from the external 
medium with the outflows that emanates from the protostar  has been 
proposed to explain  the formation of 
Herbig-Haro (HH) objects. The dynamical properties of the interaction
has been simulated by \cite{Raga_etal03}.
In high mass microquasars (HMMQ), the interaction of clumps from the 
companion stellar wind (\cite{Owocki_06}) with the jets of the compact 
object can produce  significant amount of $\gamma$ rays in the form of 
flaring and steady emission (\cite{Araudo_etal09}).
In the case where only few clumps interact with the jet, the produced
emission is sporadic and can 
explain the GeV flares detected from some $\gamma$-ray binaries 
(e.g. \cite{Abdo}).   
%

At extragalactic scales, 
active galaxies are composed by an accreting supermassive black hole (SMBH) at
the center and powerfull jets (\cite{Begelman}). 
Surrounding the SMBH there are a population 
of clouds (\cite{Krolik}) that moves at velocities $> 1000$~km~s$^{-1}$
forming the called broad line region (BLR). In the present contribution, we 
illustrate the most important dynamical and radiative consequences of the 
interaction of clouds from the BLR with the base of AGN jets.

\section{The jet-cloud interaction}

We adopt clouds with density $n_{\rm c}=10^{10}$~cm$^{-3}$, size 
$R_{\rm c}=10^{13}$~cm, and velocity 
$v_{\rm c}=10^9$~cm~s$^{-1}$. The jet Lorentz
factor is fixed to $\Gamma=10$, implying $v_{\rm j}\approx c$, and the
 radius/height
relation is fixed to $R_{\rm j}=0.1\,z$. The jet density $n_{\rm j}$ in the
laboratoty RF can be estimated as
$n_{\rm j} = L_{\rm j}/((\Gamma-1)\,m_{\rm p}\,c^3 \sigma_{\rm j})$, where
$\sigma_{\rm j} = \pi R_{\rm j}^2$ and 
$L_{\rm j}$ is the kinetic power of the matter-dominated jet. 

For a completelly penetration of a cloud into the jet,
the ram pressure of the later should not destroy the former before the cloud 
has fully entered into the jet. This means that
the time required by the cloud to penetrate into the jet,
$t_{\rm c}\sim 2 R_{\rm c}/v_{\rm c} = 2\times 10^4$~s, 
should be shorter than the cloud lifetime inside the jet.
To estimate this cloud lifetime, we first
compute the time required by the shock in the cloud to cross it 
($t_{\rm cs}$).  The velocity of this shock is $v_{\rm cs} \sim \chi^{-1/2}\, c$,
where $\chi = n_{\rm c}/n_{\rm j}(\Gamma-1)$, is derived by ensuring that the 
jet and the cloud shock ram pressures are equal. 
This yields a cloud shocking time of
\begin{equation} 
t_{\rm cs} \sim \frac{2R_{\rm c}}{v_{\rm cs}}\simeq
7\times10^4 \left(\frac{R_{\rm c}}{10^{13}\, \rm{cm}} \right)\,
\left(\frac{n_{\rm c}}{10^{10}\,\rm{cm^{-3}}}\right)^{1/2} 
\left(\frac{z}{10^{16}\,{\rm cm}}\right)\left(\frac{L_{\rm j}}{10^{44}\,\rm{erg\,
s^{-1}}}\right)^{-1/2}\,{\rm s}\,.
\end{equation} 
Rayleigh-Taylor (RT) and Kelvin-Helmholtz (KH) instabilities produced by the 
interaction with the jet material will affect the cloud. 
The timescale in wich these instabilities grows up to a scale lenght 
$\sim R_{\rm c}$ is $\sim t_{\rm cs}$. For this reason, we take $t_{\rm cs}$
as the characteristic timescale of our study.  
Therefore, for a  penetration time ($t_{\rm c}$) at least as short 
as $\sim t_{\rm cs}$, the
cloud will remain an effective obstacle for the jet
flow. Setting $t_{\rm c}\sim t_{\rm cs}$, we obtain the minimum value for the
interaction height $z_{\rm int}$, giving
\begin{equation}
z_{\rm int}^{\rm min}  \approx  2.5\times10^{15}
\left(\frac{v_{\rm c}}{10^9\,\rm{cm\,s^{-1}}}\right)^{-1}
\left(\frac{n_{\rm c}}{10^{10}\,\rm{cm^{-3}}}\right)^{-1/2} 
\left(\frac{L_{\rm j}}{10^{44}\,\rm{erg\, s^{-1}}}\right)^{1/2} \,\rm{cm}.
\end{equation}  
For  $z_{\rm int} > z_{\rm int}^{\rm min}$, the jet 
crossing time $t_{\rm j}$, characterized by
$t_{\rm j}\sim 2 R_{\rm j}/v_{\rm c} \sim 5\times10^5$s, is larger than
$t_{\rm cs}$. Once the cloud is inside the jet, a bow shock in the later 
is formed  on a time $t_{\rm bs} \sim Z/v_{\rm j} \sim 10^2$~s at a distace 
$Z \sim 0.3R_{\rm c}$~cm from the cloud. In this bow shock, particles can 
be accelerated up to relativistic energies more efficiently than in the 
cloud shock ($v_{\rm bs}\gg v_{\rm cs}$).

\section{Non-thermal  emission}

We focus here on the particle acceleration in the bow shock.
The non-thermal luminosity of particles accelerated in the bow shock 
can be estimated as a fraction $\eta$ of the bow shock luminosity: 
$L_{\rm nt} \sim \eta_{\rm nt}\,L_{\rm bs}$, where 
$L_{\rm bs} \sim (\sigma_{\rm c}/\sigma_{\rm j})\,L_{\rm j}$ and
$\sigma_{\rm c} = \pi R_{\rm c}^2$. The accelerator/emitter magnetic field 
in the bow-shock RF ($B$) 
can be determined by relating 
$U_{\rm B} = \eta_{\rm B} U_{\rm nt}$,
where $U_{\rm B}=B^2/8\pi$ and $U_{\rm nt}=L_{\rm nt}/(\sigma_{\rm c}c)$ 
are the magnetic and the non-thermal energy densities,
respectively. 
Fixing $\eta_{\rm nt} = 0.1$ and $\eta_{\rm B} = 0.01$, we obtain 
$L_{\rm nt} \sim 4\times10^{39}(L_{\rm j}/10^{44}\rm{erg\,s^{-1}})$~erg~s$^{-1}$
and $B \sim 10$~G.
Regarding the acceleration mechanism, we adopt the following prescription 
for the acceleration rate: $\dot{E}_{\rm acc}=0.1\,q\,B\,c$. 

We assume that
electrons and protons are injected into the bow shock region following a 
power law in energy of index 2.2 with an exponential
cutoff at the maximum electron energy.  The injection luminosity is
$L_{\rm nt}$. 
Particles are affected by different losses that balance the energy gain from
acceleration. The escape time downstream from the
relativistic bow shock consider advection
($t_{\rm adv}\sim 3\,R_{\rm c}/c \sim 10^3$~s) and diffusion
($t_{\rm diff}=3\,Z^2/2\,r_{\rm g}\,c$) timescales. In the later,
$r_{\rm g}$ is the particle gyroradius and the Bohm regime has been considered.

\begin{figure}
\begin{center}
\label{fig_1}
\includegraphics[angle=270, width=0.45\textwidth]{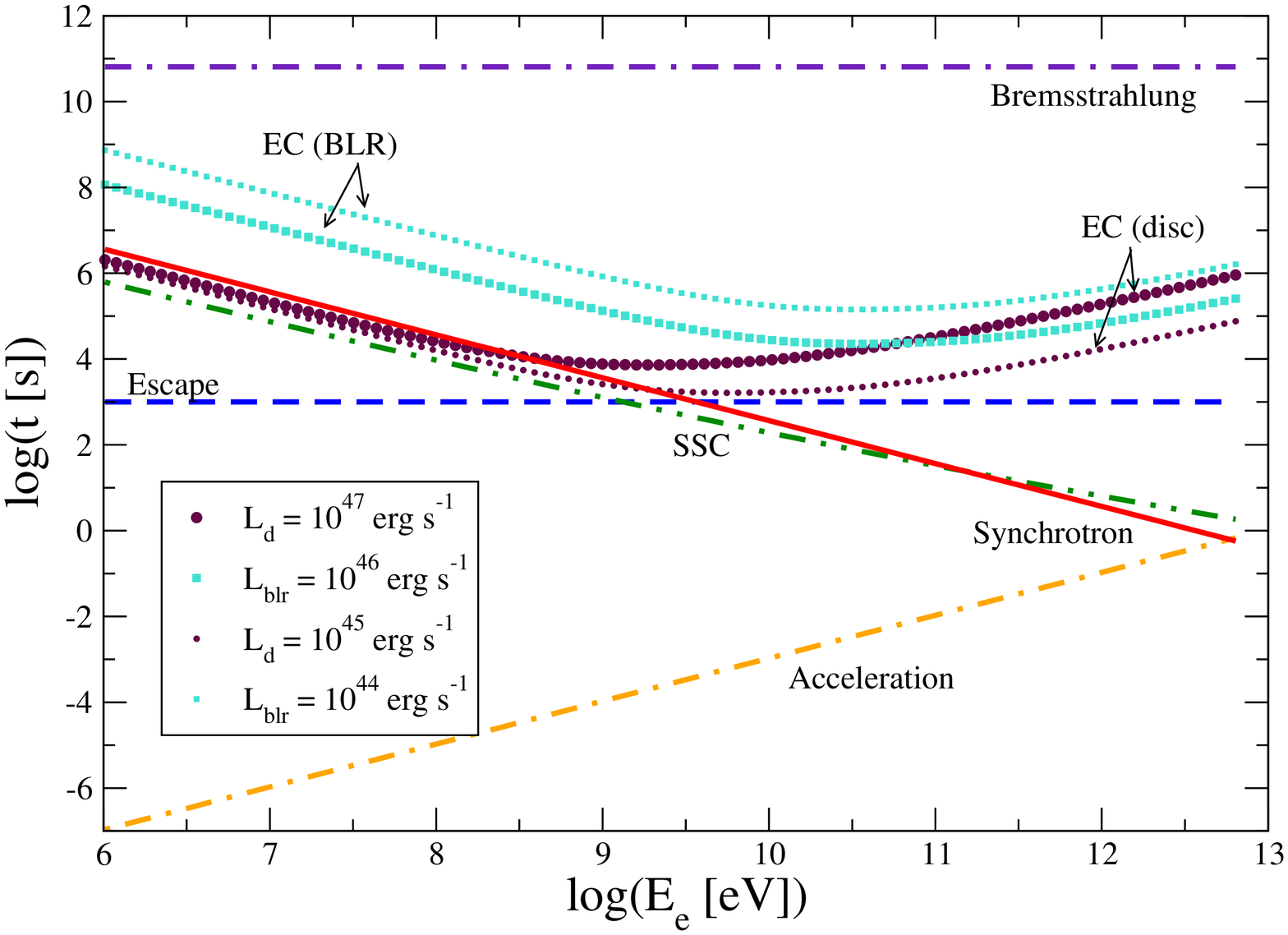}
\includegraphics[angle=270, width=0.45\textwidth]{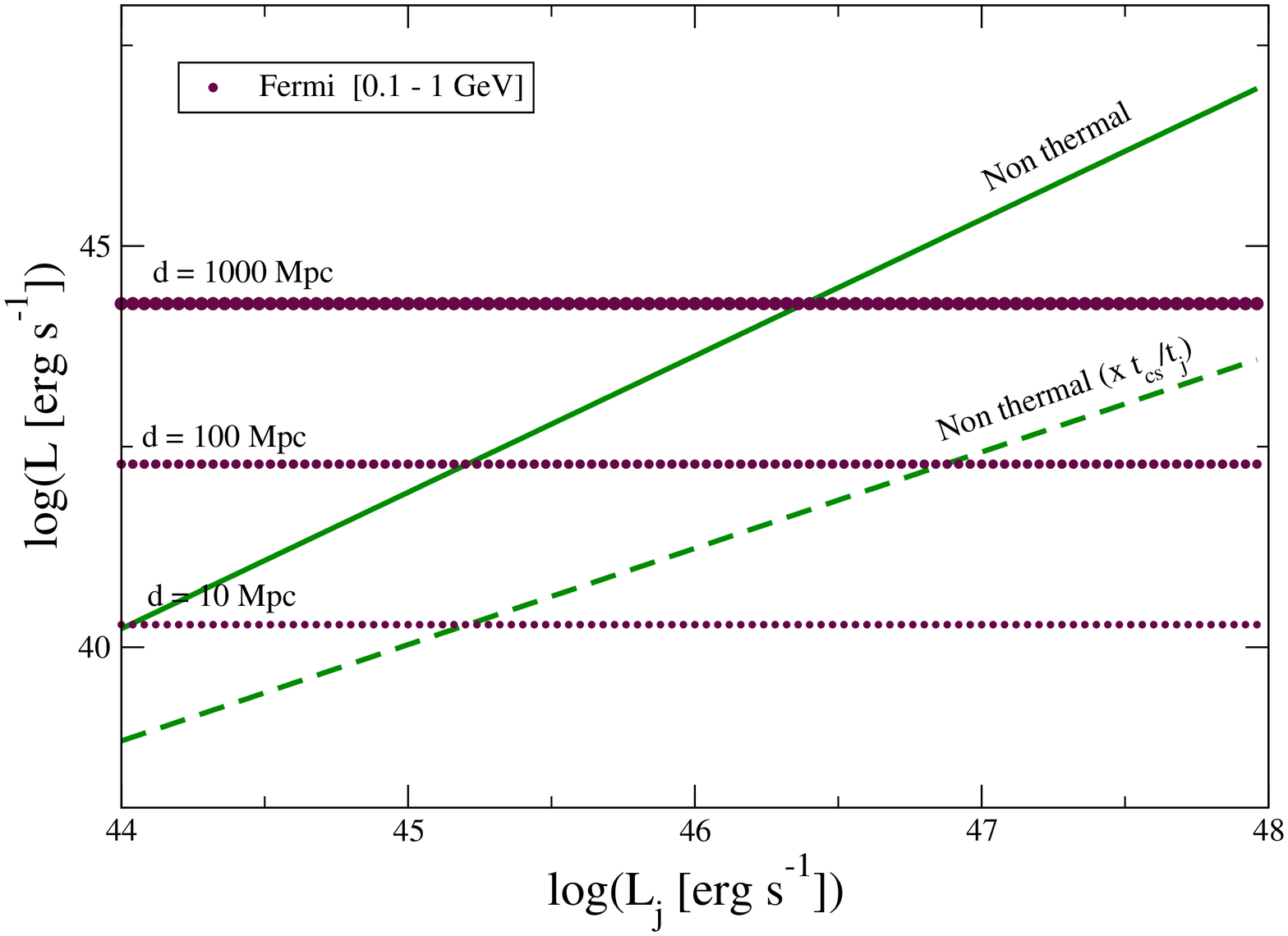}
\caption{Left: acceleration gain, escape,
and cooling lepton timescales are plotted. 
SSC is plotted when the steady state is reached
and EC for the both BLR and disc photon fields are shown for the 
conditions of faint (BLR: $10^{44}$; disc: $10^{45}$~erg~s$^{-1}$) and 
bright sources (BLR: $10^{46}$; disc: $10^{47}$~erg~s$^{-1}$). 
Synchrotron and relativistic bremsstrahlung are also plotted. 
Right: Upper limits to the $\gamma$-ray luminosity produced by 
$N_{\rm c}^{\rm j}$ clouds inside the jet as a function of
$L_{\rm j}$ in FR~II sources. Two cases are plotted, one assuming that 
clouds cross the jet without disruption (green solid lines), and one
in which the clouds are destroyed in a time as short as $t_{\rm cs}$
(green dashed lines). 
In addition, the sensitivity levels of {\it Fermi} in
the range 0.1--1~GeV (maroon dotted lines) are plotted for three 
different distances $d=10$, 100, and 1000~Mpc.}
\label{blr}
\end{center}
\end{figure}

The most important radiative losses that affect the lepton 
inyection are synchrotron and SSC, determining a maximun energy 
$E_e^{\rm max}$ of several TeV, as is shown in Figure~\ref{fig_1} (left). In
the case of protons, they 
will not cool efficiently in the bow shock via $pp$ interactions and
the maximum energy is constrained by equating the acceleration and 
diffusion timescales, given $E_p^{\rm max} \sim 
5\times 10^3\,(B/10\,{\rm G})(R_{\rm c}/10^{13}\,{\rm cm})$~TeV. Then, protons 
with energies $> 0.4E_p^{\rm max}$ can reach the cloud by diffusion and radiate
via $pp$ more efficiently than in the jet 
($n_{\rm c} \gg n_{\rm j}(z_{\rm int})$).

\section{Many clouds interacting with the jets}\label{Many-clouds}

Clouds fill the BLR, and many of them can be simultaneously
inside the jet at different $z$, each of them producing non-thermal
radiation. Therefore, the total luminosity can be much larger than that
produced by just one interaction, which is $\sim L_{\rm nt}$.  The
number of clouds within the jets, at 
$z_{\rm int}^{\rm min} \le z \le R_{\rm blr}$, can be
computed from the jet ($V_{\rm j}$) and
cloud ($V_{\rm c}$) volumes, resulting in
\begin{equation}
\label{N_clouds}
N_{\rm c}^{\rm j} = 2\, f\, \frac{V_{\rm j}}{V_{\rm c}}\sim
9\left(\frac{L_{\rm j}}{10^{44}\,\rm{erg s^{-1}}}\right)^{2}
\left(\frac{R_{\rm c}}{10^{13}\,\rm{cm}}\right)^{-3},
\end{equation}
where the factor 2 accounts for the two jets and 
$f \sim 10^{-6}$ is the filling factor of clouds in the whole
BLR (Dietrich et al. 1999). In reality, $N_{\rm c}^{\rm j}$ 
is correct if one  neglects
that the cloud disrupts and fragments, and eventually dilutes inside
the jet. For instance, Klein et al. (1994) estimated a shocked cloud
lifetime of several $t_{\rm cs}$, and Shin et al. (2008) found that
even a weak magnetic field in the cloud can significantly increase its
lifetime. Finally, even under cloud fragmentation, strong bow shocks
can form around the cloud fragments before these have accelerated
close to $v_{\rm j}$. All this makes the real number of interacting
clouds inside the jet difficult to estimate, but it should be between
$(t_{\rm cs}/t_{\rm j})\,N_{\rm c}^{\rm j}$ and $N_{\rm c}^{\rm j}$. 

The presence of many clouds inside the jet implies that the total 
non-thermal luminosity available in the BLR-jet intersection region is
\begin{equation}
\label{Lrad_tot}
L_{\rm nt}^{\rm tot} \sim 2\, \int_{z_{\rm int}^{\rm min}}^{R_{\rm blr}} 
\frac{{\rm d}N_{\rm c}^{\rm j}}{{\rm d}z} L_{\rm nt}(z)\, {\rm d}z 
\sim 2\times10^{40} \left(\frac{\eta_{\rm nt}}{0.1}\right) 
\left(\frac{R_{\rm c}}{10^{13}\,\rm{cm}}\right)^{-1}
\left(\frac{L_{\rm j}}{10^{44}\,\rm{erg\,s^{-1}}}\right)^{1.7},
\end{equation} 
where ${\rm d}N_{\rm c}^{\rm j}/{\rm d} z$ is the number of clouds
located in a jet volume  ${\rm d}V_{\rm j}=\pi\,(0.1z)^2\,{\rm d}z$. 
In both Eqs.~(\ref{N_clouds}) and (\ref{Lrad_tot}), 
$L_{\rm blr}$
has been fixed to $0.1\,L_{\rm j}$, as approximately found in FR~II
galaxies, and $R_{\rm blr}$ has been derived from \cite{Kaspi_etal02}. 

In Fig.~\ref{fig_1} (right), we show estimates of the $\gamma$-ray luminosity
when many clouds interact simultanously with the jet. For this, we
have followed a simple approach assuming that most of the non-thermal
luminosity is converted into $\gamma$ rays. 
This will be the case as long as the
escape and synchrotron cooling time are longer than the IC cooling
time (EC+SSC) at the highest electron energies. Given the little
information about the BLR available for FR~I sources, we do not
specifically consider these sources here. 
In \cite{Araudo_etal10}, we present more detailed calculations 
by applying the model presented in the present contribution to two
characteristic sources, Cen~A (FR~I, one interaction) and 3C~273 
(FR~II, many interactions).

\section{Discussion}

We have studied the interaction of clouds from the BLR with the base of 
jets in AGNs.
For very nearby sources, such as Cen~A, the interaction of large clouds
with jets may be detectable as a flaring event, although the number of
these large clouds and thereby the duty cycle of the flares are difficult
to estimate. Given the weak external photon fields in these sources,
VHE photons can escape without experiencing significant
absorption. Therefore, jet-cloud interactions in nearby FR~I may be
detectable in both the HE and the VHE range as flares with timescales
of about one day.  

In FR~II sources, many BLR clouds could interact simultaneously with
the jet. The number of clouds depends strongly on the cloud lifetime
inside the jet, which could be of the order of several $t_{\rm cs}$. 
Nevertheless, we note that after cloud fragmentation
many bow shocks may still form and efficiently accelerate particles if
these fragments move more slowly than the jet. Since FR~II sources are
expected to exhibit high accretion rates, radiation above 1~GeV
produced in the jet base can be strongly attenuated by the dense
disc and the BLR photon fields, although $\gamma$ rays below 1~GeV should
not be affected significantly. Since jet-cloud emission should be
rather isotropic, it would be masked by jet beamed emission in blazar
sources, although since powerful/nearby FR~II jets do not display
significant beaming, these objects may emit $\gamma$ rays from
jet-cloud interactions.  As shown in Fig.~\ref{fig_1} (right), close
and powerful sources could be detectable by deep enough observations
of {\it Fermi}.

\end{document}